# Unlocking the Potential of Digital Pathology: Novel Baselines for Compression

Maximilian Fischer[1,2,4,12], Peter Neher[1,2,10,11], Peter Sch¨uffler[6,14], Sebastian Ziegler[1,15], Shuhan Xiao[1,3], Robin Peretzke[1,3], David Clunie[13], Constantin Ulrich[1,4,11], Michael Baumgartner[1,3], Alexander Muckenhuber[6], Silvia Dias Almeida[1,4], Michael G¨otz[1,5], Jens Kleesiek[8,9], Marco Nolden[1,10,12], Rickmer Braren[7], and Klaus Maier-Hein[1,3,4,10,11]

1 Institute Division of Medical Image Computing, German Cancer Research Center (DKFZ), Heidelberg, Germany
2 German Cancer Consortium (DKTK), partner site Heidelberg
3 Faculty of Mathematics and Computer Science, Heidelberg University, Heidelberg, Germany
4 Medical Faculty, Heidelberg University, Heidelberg, Germany
5 Clinic of Diagnostics and Interventional Radiology, Section Experimental Radiology, Ulm University Medical Centre, Ulm, Germany
6 Institute of Pathology, TUM School of Medicine and Health, Technical University of Munich, Munich, Germany
7 Department of Diagnostic and Interventional Radiology, Faculty of Medicine, Technical University of Munich, Munich, Germany
8 Institute for AI in Medicine (IKIM), University Medicine Essen, Essen, Germany
9 German Cancer Consortium (DKTK), partner site Essen
10 Pattern Analysis and Learning Group, Department of Radiation Oncology, Heidelberg University Hospital, Heidelberg, Germany
11 National Center for Tumor Diseases (NCT), NCT Heidelberg, a partnership between DKFZ and University Medical Center Heidelberg
12 Research Campus M2OLIE, Mannheim, Germany
13 PixelMed Publishing, Bangor, Pennsylvania USA
14 Munich Center for Machine Learning, Munich, Germany
15 Helmholtz Imaging, German Cancer Research Center, Germany
maximilian.fischer@dkfz-heidelberg.de

**Abstract.** Digital pathology offers a groundbreaking opportunity to transform clinical practice in histopathological image analysis, yet faces a significant hurdle: the substantial file sizes of pathological Whole Slide Images (WSI). While current digital pathology solutions rely on lossy JPEG compression to address this issue, lossy compression can introduce color and texture disparities, potentially impacting clinical decision-making. While prior research addresses perceptual image quality and downstream performance independently of each other, we jointly evaluate compression schemes for perceptual and downstream task quality on four different datasets. In addition, we collect an initially uncompressed dataset for an unbiased perceptual evaluation of compression schemes. Our results show that deep learning models fine-tuned for perceptual quality outperform conventional compression schemes like JPEG-XL or WebP for further compression of WSI. However, they exhibit a significant bias towards the compression artifacts present in the training data and struggle to generalize across various compression schemes. We introduce a novel evaluation metric based on feature similarity between original files and compressed files that aligns very well with the actual downstream performance on the compressed WSI. Our metric allows for a general and standardized evaluation of lossy compression schemes and mitigates the requirement to independently assess different downstream tasks. Our study provides novel insights for the assessment of lossy compression schemes for WSI and encourages a unified evaluation of lossy compression schemes to accelerate the clinical uptake of digital pathology.

Keywords: **Whole Slide Images, Lossy Compression, Feature Similarity**

# 1 Introduction

Despite broad availability of digital Whole Slide Imaging (WSI) Scanning systems, the clinical realization has not progressed significantly. One major burden for laboratories is the high storage costs of digital slides. The coding of WSI scans with 24 bits (bpp) per pixel (as is usual for RGB images) leads to tremendous amounts of memory for complete scans. A first mitigation strategy to address this problem is compression schemes to reduce the length of the bitstream per pixel, which are offered by all WSI vendors. When the first commercial WSI scanner was introduced in 1994 [1, 2], the JPEG algorithm [3] was the current state-of-the-art lossy compression method, and is still being used by the majority of WSI vendors to this day. However, the JPEG algorithm was initially developed 50 years ago for the compression of natural scene images with a dedicated focus on human observers. In modern pathology applications, the requirements are slightly different. Thus, neither the natural domain nor the human-centric aspect seems perfectly suited for compressing pathology slides. Especially since it has been thoroughly shown that various other compression schemes constantly outperform JPEG regarding image quality and file size reduction [4–9]. But in clinical applications, the challenges of data compression extend beyond the limitations of specific algorithms, which are primarily centered on preserving diagnostic accuracy while efficiently reducing file sizes. Moreover, verifying the presence of medical details on lossy compressed data introduces another layer of complexity and thus preference is typically given to lossless compressions in medical imaging, due to the possibility to always reconstruct the original data. However, this comes with the tradeoff of limited degrees of compression [10] and the current compromise in WSI is the slightly lossy compression with low compression ratios (e.g., JPEG with a high quality factor between 80 and 90), as seen in cohorts like Camelyon16 [11]. This setting maintains high perceptual quality with reasonable amounts of file size reduction. However, in recent years it has been shown that for many tasks, further compression schemes beyond the initial one or improved schemes enable further reduction of the still very large file sizes without affecting diagnostic accuracy [12–22]. But still the exact evaluation of such lossy compression schemes is more complex than lossless algorithms, as they neglect the requirement of reconstructing the original data and permanently alter image information. Due to its relevance, this problem has already been evaluated from various perspectives. Previous studies in the pathology domain have predominantly concentrated on assessing the impact of combined compression schemes to further compress WSI after the initial compression. These compression schemes involve either recompression using JPEG or JPEG2000 following the initial compression. While some studies, such as [13, 22], evaluate the impact of repeated compression on algorithm-based downstream tasks using JPEG, others, including [12–14, 16, 17], assess its effects using JPEG2000. In contrast to these studies, which primarily rely on computational methods, [15, 19] employed human experts for optical evaluations of repeated JPEG compression. But all of these studies utilized different datasets featuring various tissue types, thereby limiting the generalizability of their findings. Consequently, it is challenging to draw overarching conclusions regarding recompression across tissue types based solely on these evaluations. However, this situation is not unique to the pathology domain. In other fields, such as natural scene images, there has been a significant shift towards exploring fundamentally new deep learning (dl)-based compression schemes, including compression autoencoders [23]. This trend has also recently extended to the pathology field, leading to advanced compression schemes that leverage dl techniques [24–26]. For instance, in [24], an approach is presented that aims for dimensionality reduction of WSI via an autoencoder. Here the latents are stacked to form a pseudo-WSI but human perception of the recon-structed latents is not evaluated in this study, which continues to be the clinical standard, despite emerging computerized diagnosis systems. While conventional autoencoder settings are employed in this study, both [25] and [26] propose the use of a Vector-Quantized Variational Autoencoder for lossy WSI compression. [25] focuses on fine-tuning to enhance perceptual quality, while [26] prioritizes improved downstream performance. However, it is worth noting that unlike [24], these studies do not target dimensionality reduction, as it is not equivalent to lossy compression. Despite these exhaustive evaluations of lossy compression of WSI, digital pathology is currently still hampered by large file sizes. A major drawback of existing studies in the field is that the studies that evaluate recompression schemes, mostly neglect dl-based compression schemes. A joint evaluation in a standardized setting that considers perceptual and downstream evaluation is missing. Different datasets and tasks between different studies also limit the generalizability of individual findings. Another drawback is that most studies also perform their evaluations on initially JPEG compressed data, lacking assessments on initial compression schemes.

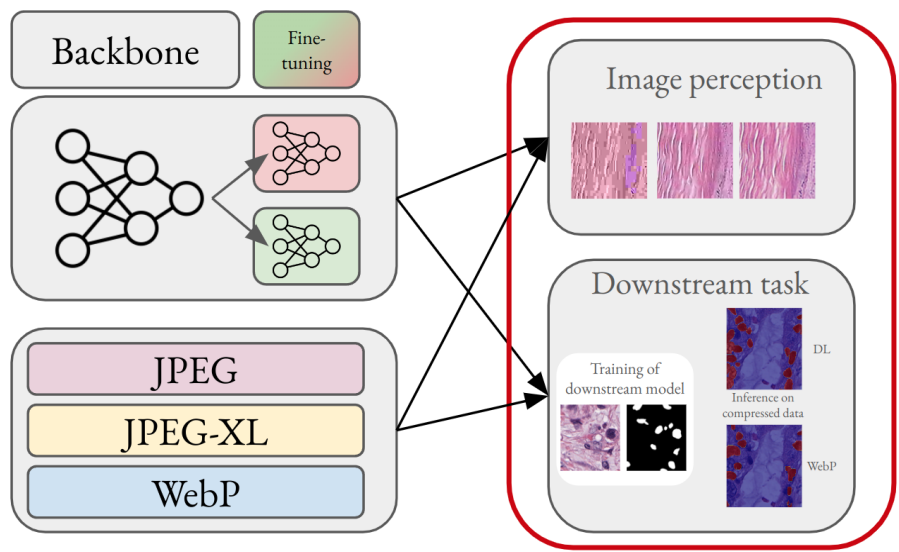

Fig. 1: Overview of our approach. We provide a full overview of the performance of linear compression schemes as well as for more recent dl-based schemes on human perception, as well as for dl-based downstream tasks.

In this study, we aim to address the shortcomings of previous research by conducting a comprehensive evaluation of compression schemes, focusing on both perceptual image quality and downstream task performance. We are the first study that systematically evaluates lossy compression schemes for perceptual quality as well as downstream tasks in digital pathology. Our approach is shown in Figure 1 and our contributions are as follows:

- Evaluation of state-of-the-art compression schemes: We provide a holistic assessment of state-of-the-art lossy compression methods, evaluating their performance during recompression with respect to perceptual image quality metrics and in downstream tasks.
- Evaluation of initial compression schemes: We extend existing work by evaluating lossy compression schemes during the initial compression of uncompressed data.
- Task agnostic downstream metric: We introduce a novel, more general metric for evaluating downstream performances across several tasks.
- Computational evaluation: We offer a comprehensive evaluation of the computational aspects of compression schemes, highlighting their direct relevance to real-time image analysis, which is closely connected to the acceptance of digital WSI systems.

Our study provides a standardized and novel evaluation of compression schemes during recompression and initial compression experiments. We show that dl-based compression schemes can be finetuned to outperform conventional compression schemes on specific datasets for both the perceptual and downstream evaluation, but also exhibit a large dataset bias and show poor results when they are applied out of domain. Our results show that the proposed task-agnostic metric for downstream performances aligns very well with the actual downstream performance. We encourage other studies to also evaluate compression schemes on task-agnostic metrics to improve comparability. Our results show that dl-based compression schemes are not yet ready to replace JPEG, but other conventional schemes like WebP show promising results to bridge the gap between JPEG and until dl compression schemes can go into clinical application. We show that one major burden for the clinical application of dl-based compression schemes are the unreasonable longer compute times compared to conventional schemes. The article is structured as follows. In section 2, a detailed overview of conventional and dl-based compression schemes is provided, where a strong focus is on the used datasets, on which the schemes are evaluated. In the next section 3, the experiments are split into two tracks: the perceptional track and the downstream track. In both tracks we compare the same compression schemes in a standardized experimental setup, but with task-specific metrics and on suitable datasets. In the result Section (section 4), we quantify the performance of each compression scheme, and compare also the amount of compute time for various schemes for en- and decoding of images. Finally, we discuss each method and provide conclusions our studies.

# 2 Study design and Benchmark methods

Image compression and especially lossy image compression schemes are developed to optimize a rate-distortion objective. This formalism describes the trade-off between the two competing costs: the bit rate used to represent the compressed data (rate) and the distortion between input and reconstruction that arises from the quantization (distortion). Quantization describes the irretrievable deletion of information, which is what lossy compression applies and what distinguishes it from lossless compression. In Our study, we evaluate lossy compression schemes for two clinically relevant tasks in digital pathology: (i) perceptional image quality for diagnostics by pathologists and (ii) downstream performance in deep-learning-based applications. For both targets, we compare 3 conventional schemes, JPEG, JPEG-XL and WebP against 3 deep-learning based lossy compression schemes in a standardized setting. The following sections introduce the set of benchmark compression schemes, that we consider in our evaluation, followed by an exhaustive description of the datasets. We also present the compared metrics in this study, where we propose a novel task-agnostic metric to determine downstream task performances on lossy compressed images.

## 2.1 Benchmark Methods and state of the art

**Conventional Compression Schemes** The most common lossy compression is the JPEG algorithm [3]. The basic algorithm replicates two principles of the human perceptual system: (1) changes in brightness are more relevant than changes in color and (2) the higher sensitivity towards low-frequency components, compared to high-frequency components [27]. The lossy JPEG algorithm applies both characteristics in turn and is often utilized as a solid baseline in compression evaluations. For a comparison with existing studies in the field, we also consider JPEG in our study. Diverged from the baseline JPEG algorithm, several derivatives like JPEG2000 emerged, which have already been assessed in various other studies. Thus we consider JPEG-XL [28] in our study, which is a more substantial improvement compared to the baseline approach. JPEG-XL, allows more flexible window sizes instead of the fixed 8x8 blocks in JPEG and specifically aims for improved web delivery and compressing high dynamic ranges in images. Another conventional compression scheme, that also aims for improved web delivery is WebP. Compared to JPEG-XL, WebP achieves higher compression ratios at similar visual quality. WebP is based on the predictive VP8 [29] block prediction approach and divides the image also into smaller segments. Here an encoder predicts redundant color information in each processed segment, based on the previously processed segment. WebP encodes only the difference between the predicted value and the actual value, which enables very high compression ratios. For our study, we compare three conventional compression schemes. We consider JPEG due to its broad distribution and as the current standard in WSI. Besides that, we compare two more recent compression schemes that are specifically tailored for web-based applications, which are closely connected to digital WSI.

| Feature | JPEG (1992, **currently applied at WSI**) | JPEG-XL (2022) | WebP (2010) |
|---|---|---|---|
| Compression Algorithm | Discrete Cosine Transform (DCT) | Advanced DCT with adaptive quantization | VP8 (Video compression technique adapted for images) |
| Block Size | Fixed 8x8 blocks | Flexibel block sizes enable multi-scale transforms | Fixed 4x4, 8x8 or 16x16 blocks |
| Color Coding | YCbCr | RGB, YUV | RGB, YUV |
| Image Reconstruction | Block-based decoding leads to visible image artifacts for high compression ratios | Adaptive DCT leads to fewer blocking artifacts | Predictive coding leads for high compression ratios to noisy reconstructions |
| File sizes/Processing Speed | Large / Fast | Small / Slow (esp. During encoding) | Small / Fast |
| Applications | Fotography / **digital Pathology** | High bit depth (native HDR support) | Optimized for web-based applications |
| Advantages | Current scanning Hardware is widely optimized for JPEG compression and natively supported by all WSI vendors, but technology-wise outdated | JPEG-XL ist the modern successor of JPEG and addresses most of the downsides of the conventional JPEG algorithm and largely improves the image quality of WSI | WebP with it's fast coding times and small file sizes is ideally suited for web based applications which are increasingly important in digital pathology |

Table 1: Comparative overview of the considered conventional compression schemes. What is considered as an advantage is highlighted by the green color, while potential disadvantages are highlighted with red.

**Neural Compression Schemes** As a more recent and flexible approach, we compare the conventional schemes with deep learning-based compression schemes. Similar to most conventional linear compression schemes, neural codecs also transfer an input image into another domain that is better suited for quantization. However, dl-based approaches are not constrained to linear transformations in their search space and can possibly find better domains for quantization than conventional approaches. Most neural image compression schemes are mainly based on compressive autoencoders (VQ-VAE) [9] and consist of three components: (i) A dimensionality reduction network $f\psi$, that learns non-linear transformations to generate the latent representation $y$ of the input data $x$. (ii) A shared distribution model between the encoder and decoder $p\psi$ that models the distribution of the latents, which is necessary for a lossless entropy-based coding algorithm for the latents. (iii) A decoder model $g\psi$ that reconstructs the quantized latents for the reconstruction of the image $\hat{x}$. This formalism is usually modeled as follows during training: First, the distortion $d$ between the original image $x$ and the reconstructed image $\hat{x}$, and second the length of the expected bitstream $r$ of the compressed output shall be minimal. For training of such models, metrics, like the Peak signal-to-noise ratio (PSNR) can be used to determine the image distortion, and the expected bitrate, is determined by Shannon [30]. In this paper, we consider three neural image compression schemes, that follow the presented bitlength-distortion optimization. As a baseline compressive autoencoder model, we use the *bmshj-factorized* from [31]. The model emerged from the natural scene image domain and is well-established in the field. The model is pre-trained on the Vimeo-90K [32] dataset, which contains a large variety of uncompressed scenes and actions, and thus no explicit knowledge about medical images is incorporated during pretraining and no bias towards any initial compression. Compress AI [33] provides implementations for this model and pretraining and we thus refer to this model in the following as *CAI*. Besides this *CAI* model, we also consider two approaches that are tailored towards the compression of pathology data [25, 26]. In contrast to the baseline model, the *SPL2* model from [25] was trained on a broad range of pathology images during training and it was shown that this model achieves state-of-the-art results for recompressing *JPEG80* data for perceptual image quality metrics. The model was additionally supervised with an additional loss term for improved realism of the reconstructed images. On the other hand, downstream performances on compressed WSI is almost equally important as perceptual quality and in [26], a method is presented that addresses this aspect. We refer to this method as *SQLC*. Here additional staining information is incorporated during the compression, which achieves state-of-the-art downstream performance on compressed WSI. An additional encoder first encodes staining information together with the RGB-image, before the main model encodes all information jointly, which is not feasible with one single encoder. For our study we selected the *CAI* approach, since this is a well-established approach from the natural scene image domain and serves as baseline deep learning compression as comparison. Additionally, we select the *SPL2* and *SQLC* models, which are both having the same architecture as the *bmshj-factorized* model. The *SPL2* is specifically finetuned towards improved perceptual image quality, whereas *SQLC* is finetuned for downstream performances and thus both models are ideally suited for our holistic evaluations.

| Feature | SQLC | SPL2 | CAI |
|---|---|---|---|
| Focus | Pathology-specific (H&E-stained WSIs). | General pathology WSIs with emphasis on perceptual fidelity. | General purpose image compression research. |
| Key Technique | Combines deconvoluted stain channels and RGB for compression. | Embedding similarity with contrastive pre-training. | Pre-trained learned codecs with modular training. |
| Optimization Goals | Stain-aware compression and classification accuracy. | Perceptual fidelity and diagnostic relevance. | General rate-distortion efficiency. |
| Metrics | MS-SSIM, AUC for tasks | LPIPS, PSNR, MS-SSIM | PSNR, MS-SSIM, bit-rate |
| Application Specifity | Tailored to H&E-stained slides | Generalizable across pathology tasks and stains. | Board applicability to image compression. |
| Strengths | Compressing downstream relevant image features. | High-quality visual reconstructions for manual analysis. | Large scale pre-training (but only natural scene images). |

Table 2: Comparative overview of the considered neural compression schemes.

## 2.2 Datasets

For our study, we collect a total of four different histopathological datasets, where we use two datasets for the perceptual evaluation (*JPEG80* and *RMS*) and two datasets for the downstream experiments (Camelyon16 and PanNuke). Each dataset is established in the field and several other studies have used one of them for their evaluations. All datasets are publicly available in order to enhance Reproducibility.

**JPEG80 Dataset** The first dataset, the *JPEG80* dataset is used to evaluate recompression capabilities of compression schemes. We refer to this dataset as *JPEG80*, since all data in this dataset is initially JPEG compressed with a quality factor of 80. This dataset is a collection of several publicly available datasets [34–36] from various tissue types and different centers that are compressed with *JPEG80*. [34–36] into one dataset, the *JPEG80* dataset, which was also used by [25] for their evaluations. In table 3, an overview of the dataset for this study is presented. We use this dataset to assess the impact of recompression on a broad range of tissue sections and staining characteristics. The dataset was initially introduced in [25] and for our evaluations, we only use the test split, which contains the same samples as in the original publication.

| Dataset | BreaKHis | Colon1 | Colon2 |
|---|---|---|---|
| Source | [34] | [35] | [36] |
| Tissue | Breast | Colon | Colon |
| Images | 400 | 100 | 2 |
| Sample Size | 700x460 | 768x768 | 5000x5000 |
| Tile Size | 224x224 | 224x224 | 224x224 |
| Tiles | 2000 | 2000 | 1000 |
| | 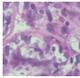 | 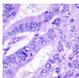 | 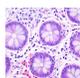 |

Table 3: The used dataset is equivalent to the one used in [25]. The dataset contains various tissue types to address a broad spectrum of clinical applications.

**Uncompressed RMS data** Data without any initial lossy compression are sparse, due to the high storage costs and thus no studies have investigated the impact of recent lossy compression schemes on lossless compressed pathology Whole Slide Images. With the Rhabdomyosarcoma (RMS) images [37] that are now available on Imaging Data Commons (IDC) [38], one of the few large-scale WSI cohorts is now publicly available that has not been previously lossy compressed. We refer to this dataset as the *RMSdata* from which we collected 100 randomly selected subjects. We performed foreground detection by simple thresholding and extracted tiles with the size of 224 pixels from the highest magnification level on random positions without any overlap between tiles. In total, we extracted 1000 tiles per subject, which results in 100,000 uncompressed tiles. An overview of the data set as well as some sample images can be found in Figure 2.

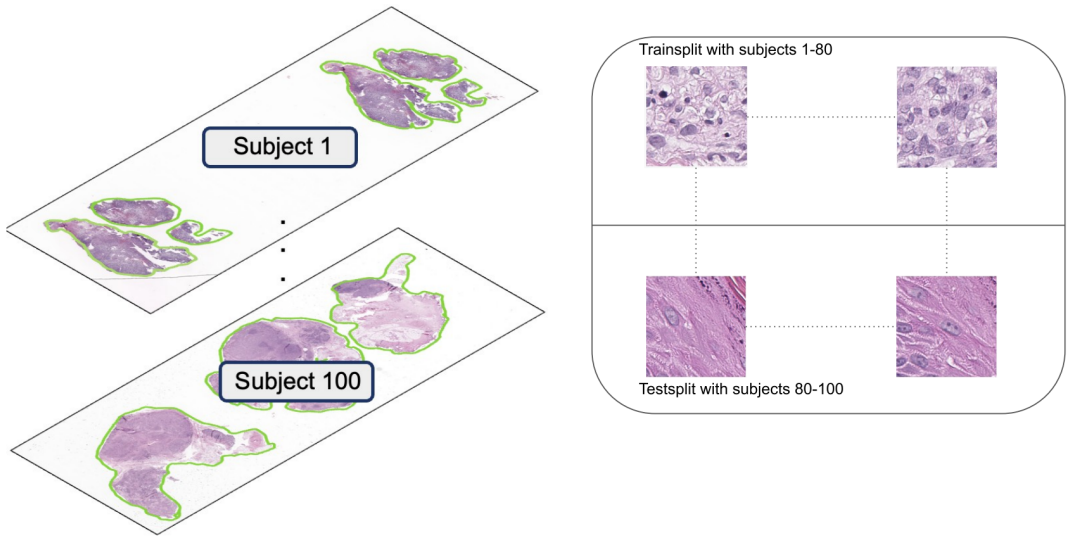

Fig. 2: Overview of the RMS dataset. The dataset was acquired without lossy compression during acquisition and is thus ideally suited to evaluate compression schemes for the initial compression. During five-fold cross-validation, we split the data on a patient label into folds.

**Camelyon 16 data** While image quality tasks can be evaluated unsupervised, downstream evaluations are based on annotated images. Such datasets are less often published and one of the few annotated datasets is the Camelyon16 [11] WSI dataset. The same dataset was also used in [26] for their evaluations. This dataset is well-established in the histopathology domain and it contains lymphnode WSI and the 270 training and 130 test subjects are divided on a subject level into the classes metastasis and non-metastasis with the same splits as in [26]. Ground truth annotations for the subjects are available on a pixel level. From the WSI files, we extract a maximum of 500 non-overlapping patches per class and patient at random positions on the slide. One patient can only belong to one class (i.e. we do not sample non-tumor patches from tumor subjects) and we set the patch size s = 224. All sampled metastatic tiles must at least contain 30% tumor tissue and all patches from all classes are sampled at the highest available magnification. From the 110 tumor and 160 non-tumor slides in the train set, this sampling yields 28735 tumor and 79418 non-tumor tiles, while for the test set 48 tumor and 82 non-tumor slides resulted in 11097 and 40500 tiles.

**PanNuke data** As an additional downstream dataset, we consider PanNuke [39, 40]. This dataset is one of the largest quality-controlled datasets for nucleus segmentation that is publicly available. The dataset contains samples from 19 tissue sections and different classes (Neoplastic cells, Inflammatory, Connective/Soft tissue cells, Dead Cells, Epithelial, Background) of the annotated nuclei. The ground truth instance segmentations were generated in a semi-automatic way, where the segmentations were verified by clinical professionals. In total, the dataset consists of patches from more than 20k WSI samples from various data sources. Each patch has 256x256 pixels and the dataset is provided as precomputed patches with corresponding segmentation masks. As well as the Camelyon16 dataset, also PanNuke was initially *JPEG80* compressed.

# 3 Experiments

The evaluation of Perceptual Image Quality and downstream performance requires distinct measures and datasets. As a result, we already divided our datasets according to different evaluations. We follow this distinction throughout the study and divide the experiments in a perceptual quality track and a downstream track. In both branches, we follow a standardized evaluation approach and provide an exhaustive comparison of compression schemes across both tracks.

## 3.1 Perceptual Image Quality Track

**Perceptual Metrics** Comparing the performance of the methods that are introduced in section 2 numerically is challenging. Image artifacts are primarily subject to the individual preferences of human observers and their correlation with numerical metrics is a field of active research. Currently, there is no metric available that can reliably predict human judgments of image quality in compressed images [41]. Nevertheless, certain metrics are established in the field of lossy compression, either for evaluating compression algorithms or as metrics during training of dl-based compression schemes. For our evaluation, we consider PSNR, MS-SSIM [42], and LPIPS[43]. The most common metric is the Peak-Signal-to-Noise-Ratio (PSNR), which is derived from the Mean Squared Error (MSE). The metric compares the true pixel values between two images pixel for pixel and due to its simplicity, this metric is the baseline evaluation to assess distortions and thus also quality. However, one of the biggest disadvantages of this metric is the omission of neighborhood relations between pixels when calculating the MSE error. But images are spatially arranged data and strongly depend on neighborhood relations between pixels, so recent work has also shown that optimizing VQ-VAE models directly on the MSE error leads to noisy reconstructions and insufficient image quality. Nevertheless, the MSE error is intuitive to interpret and commonly used and thus we consider the PSNR metric for our study as well. A more recent metric, that partially addresses the downsides of the PSNR metric is the Multi-Scale Structural Similarity Index Measure (MS-SSIM). The MS-SSIM measure is based on the SSIM metric and is also its most commonly used extension. While the PSNR metric is calculated on a pixel level, SSIM is computed for small image patches, that also consider relationships between neighboring pixels. The evaluation of the SSIM at multiple resolutions in an image yields the MS-SSIM metric, which has been shown to correlate much better with human perception. But again, also this metric does not exhaustively quantize each facette of perceptual quality and it has been shown that this metric as well as PSNR tends to fail on data that was created by generative approaches like GAN or Diffusion models [41, 44, 4]. This shortcoming is addressed by the Learned Perceptual Image Patch Similarity (LPIPS) [43] metric. In contrast to other metrics, which measure the distance between two images in the RGB-image domain, LPIPS evaluates the image quality via the $\ell 2$ distance between deep embeddings across all layers of a pre-trained network between two images. In their original publication, the authors of the LPIPS metric suggest using an ImageNet pre-trained VGG neural network to generate the embeddings from the input images. It has been shown that this metric leads to more realistic reconstructions of dl-based compression schemes. For our study, we consider all three metrics, since each one captures individual and equally important image features. Besides low distortions in the reconstructed images, which is evaluated by PSNR and MS-SSIM, we also evaluate the realism of the reconstructed images, which is evaluated by LPIPS.

**Perceptual track experimental setup** In the perceptual image quality track of this study, we want to address the following research questions: (i) Which compression scheme should be ideally applied to already acquired data in order to reduce storage costs for laboratories in the short and mid-term and (ii) which compression scheme should be deployed in the long-term as the successor of JPEG during image acquisition. To address these questions we perform 3 types of experiments. In the first experiment (Recompressing *JPEG80* data), we apply all compression schemes on precompressed data. Here we use

*JPEG80* dataset to apply the 6 considered compression schemes within the range of an effective resulting bitrate between 0 and 1.75 after the combined compression. For the implementation of the compression schemes, we use Ubuntu Terminal applications for the compression of [JPEG17](#), [JPEG-XL18](#) and [WebP19](#) and the deep learning models are implemented via checkpoints, that are provided from the original publications from the respective method. For the dl-models we freeze all weights and perform no finetuning. For the experiments, we use the validation dataset that was introduced in 2.2 and report the perceptual image metrics. Here we average across the whole dataset and report the mean and standard deviation. We use only the test split of this dataset for our evaluations, which is the same split as in [25]. In the second experiment (Compressing *RMS* data) we want to address what compression scheme is ideally suited for the initial compression. Thus, we compare the compression schemes on the uncompressed *RMS* data. We use the same implementations as previously described to generate the compressed data. Again, we measure the perceptual quality scores and determine the mean and standard deviation across the whole test set. As an additional experiment here, we also use the *SPL2* method and also retrain it on the uncompressed *RMS* data. We train the method accordingly to the original publication [25] and we perform five-fold cross-validation where we use 80% of the available subjects in the train set of each fold and 20% as test subjects. In the third experiment (Finetuning on *RMS* data) in this track, we also finetune a recompression scheme on initially uncompressed data. Most of the current studies have evaluated their approaches for further compression already on initially JPEG compressed data and thus, only limited evaluations of the general recompression capabilities are possible. To illustrate this bias, we compress this dataset with JPEG, JPEG-XL and WebP with a compression ratio that is comparable with *JPEG80*. Here we focus on conventional compression schemes during the initial compression. Subsequently the current state-of-the-art approach for further compressing WSI images *SPL2* from [25] is applied on top of the JPEG, JPEG-XL, and WebP initial compression. For this setup, we take the trained model from [25] and finetune it 10 more epochs for the specific compression artifacts of each scheme. We use a learning rate of 1e-5 for each finetuning and we train one model on top of each compression scheme. Again we use five-fold cross-validation, as described in the experiment of compressing RMS data.

## 3.2 Downstream Image Quality Track

Downstream Metrics In contrast to perceptual image quality metrics, downstream task metrics depend on the specific task that is evaluated. In our study, we extend the evaluation from *SQLC* by a semantic segmentation task, which requires to use the Dice score as a downstream metric. A common shortcoming of such evaluations is that each task requires an individual metric and evaluations across various tasks are difficult. We mitigate this shortcoming and the need for tedious implementations of single downstream tasks with a more abstract and task-agnostic view of downstream performances. Our approach is inspired by the domain of natural scene images [43] and is based on the assumption that independently from the specific task that has to be conducted, it is assumed that, the upper baseline is the original image data. For an image that is very similar to the original image, which can be a compressed version of the original image with few image artifacts, it is also assumed that the downstream performance on that image is comparable to the performance on the original data. In our approach, we measure the stability of image-relevant features during compression, similar to [43] also in the latent domain, and use it as task agnostic downstream evaluation. As a feature extractor, we use the model from [45], which was one of the first foundation models in digital pathology. The model has been trained on a wide range of pathological images and shows superior performance in various downstream tasks compared to ImageNet pre-trained models and is thus ideally suited for feature extraction. For our evaluation, we use the ResNet18 foundation model as feature extractor, where we consider the features after the (1.) first convolutional layer and the first set of residual blocks, (2.) the second set of residual blocks, (3.) the third set of residual blocks, (4.) the fourth set of residual blocks, (5.) after the average pooling layer, before flattening, and (6.) after the final fully connected layer. We compute the features for the original image and its compressed counterpart, where the cosine similarity between both sets of features after flattening is given by:

$$s_{cos}(x, y) = \frac{\sum_{d=1}^{D} x_d \cdot y_d}{\sqrt{\sum_{d=1}^{D} x^2{}_d} \cdot \sqrt{\sum_{d=1}^{D} y^2{}_d}} \qquad (1)$$

where x represents the features that are extracted from the original image and y the features from its compressed counterpart. What is usually desired after compression, is that the reconstructed image is as similar to the original image as possible. Thus embeddings of images should ideally point towards the same direction in the feature space, which is captured by the cosine similarity.

**Downstream track experimental setup** It has been shown that image features that are relevant for humans to analyze an image are not necessarily correlating with the features that are important for machines during image analysis [46]. The superiority of dl-based compression schemes for downstream tasks has already been shown in [26] for the example of a classification task and in this paper we extend existing evaluations and address the following questions: (i) How effectively do existing evaluations generalize across different computational pathology tasks and (ii) is there a more generalized approach to evaluate downstream performances task-agnostic? In the first experiment (Segmentation task) we implement a nuclei segmentation task, which is a very common task in digital pathology [13, 11]. In this experiment, we evaluate the degree of lossy compression in the test data, without affecting the performance of the segmentation model, which was trained on the original data. For the training of a segmentation framework, we use the nnU-Net [47]. We train the model on the original PanNuke dataset with the default nnU-Net training schemes, we employ three-fold cross-validation, where we use the provided folds from the dataset and measure the Dice metric. As compression schemes, we consider again the same set of 6 compression schemes and compress each validation fold. In the second experiment (Task agnostic metric), we assess the task-agnostic downstream evaluation. In our experimental setup, we combine the test sets from the PanNuke and Camelyon datasets into one combined dataset and create compressed versions of the test set with all 6 compared compression schemes to a bit rate of 0.5. For evaluation, we form image pairs of the original image and its compressed version. For each image within one pair, we extract features with the foundation model and measure the distance between both sets of features.

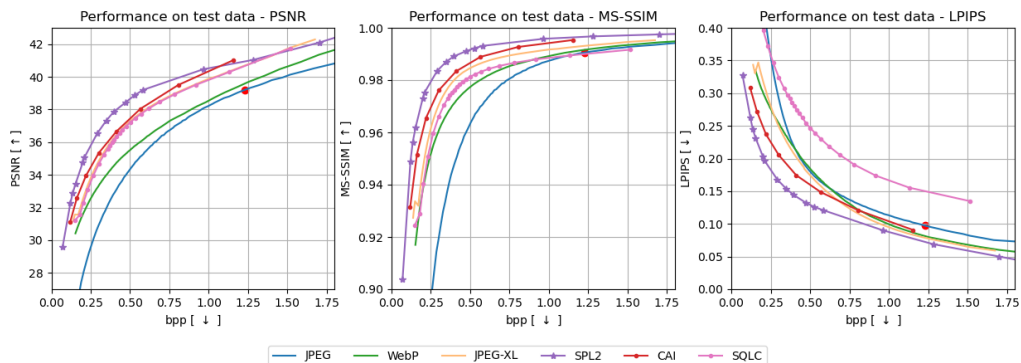

Fig. 3: Rate-distortion and -perception curves on the test set for various metrics. Arrows (↑), (↓) indicate if high or low scores are preferable. The bits-per-pixel (bpp) determines the level of compression.

In our experiments, we evaluate both perspectives: perceptual image quality and downstream performance. Visual inspection of pathology slides by pathologists remains the clinical standard, despite emerging deep-learning based automated analyses. As outlined, image compression schemes can degrade the perceptual quality and potentially degrade the pathologists ability to identify critical image features. Quantitative metrics like MS-SSIM or LPIPS provide objective measures to evaluate the perceptual quality. Although qualitative assessments by pathologists can provide additional insights, it is always suspect to personal preferences and thus only partially suited as general evaluation metric. In addition to the human-centric applications, deep-learning based automated classification, segmentation or biomarker detection has revolutionized the field of digital pathology, leading to multiple pathological foundation models [50,51,52]. Evaluating compression in the context of deep-learning workflows ensures that automated systems maintain robust performance and avoids unanticipated failures during real-world deployment.

## 3.3 Compression Time Evaluation

A key aspect that practical WSI processing algorithms must fulfill is fast and efficient processing. The processing of large amounts of data, which is already implicit in large-scale WSI scans, requires algorithms that are as fast as possible in order to be reasonably practicable. To measure the encoding times, we write the files to the respective file format for the conventional schemes on the disk (.jpeg, .jxl, .webp). For the dl-based compression schemes, we write the compressed files as binary file. For the evaluation of the decoding performance, we decode each encoded image into the RAM of a Ubuntu 20.04 workstation, which is equipped with 64 GBs of RAM and an AMD Ryzen 9 3900X 12-Core Processor. We perform this evaluation on the *RMS* dataset. For this experiment, we encode single tiles, but the *RMS* dataset roughly is the size of one complete WSI and in practice, WSI are also encoded tile-wise. The results from this experiment can thus be compared with the performance on one WSI file. In this experiment, we measure the time that is required to en- and decode the whole *RMS* dataset with all 100,000 tiles of size 224 pixels each.

## 4 Results

## 4.1 Perceptual Image Quality

Recompressing *JPEG80* data In Figure 3, we show various perceptual image quality metrics for different compression schemes. The x-axis in the plots shows the bit rate of the images and along the y-axis, we report the quality metric. Each plot shows the mean and the standard deviation as a shaded area. Please note that the standard deviation was throughout our experiments very small and in the range of ±0.001. The Figure shows that the perceptually finetuned model from [25] outperforms all other compression schemes, including the downstream fine-tuned model, for all compared metrics. Compressing *RMS* data Similar to the previous section, we again report the rate-distortion performance for the perceptual metrics in Figure 4a. The bitrate of the JPEG compression with a quality factor of 80 as reference is marked with a red dot in all three plots. The Figure shows, that large portions of perceptual image quality are already lost when JPEG is applied for the initial compression and other compression schemes achieve higher perceptual image quality compared to JPEG for the same compression ratio. Finetuning on *RMS* data As an extension, we investigate the impact of the best-performing dl-based compression scheme, when it is applied on top of conventional compression schemes. This is shown in Figure 4b. Here, the *RMS* data is compressed with JPEG, WebP, and JPEG-XL to a comparable bpp of *JPEG80* and we apply the current state-of-the-art compression model *SPL2* on top. In Figure 4b, each curve refers to one of the conventional schemes, where *SPL2* is applied on top. The figure shows that the combination of WebP and *SPL2* performs best in all metrics for a recompression task. For visualization purposes, we only show MS-SSIM and LPIPS, but the same trend also manifests in PSNR.

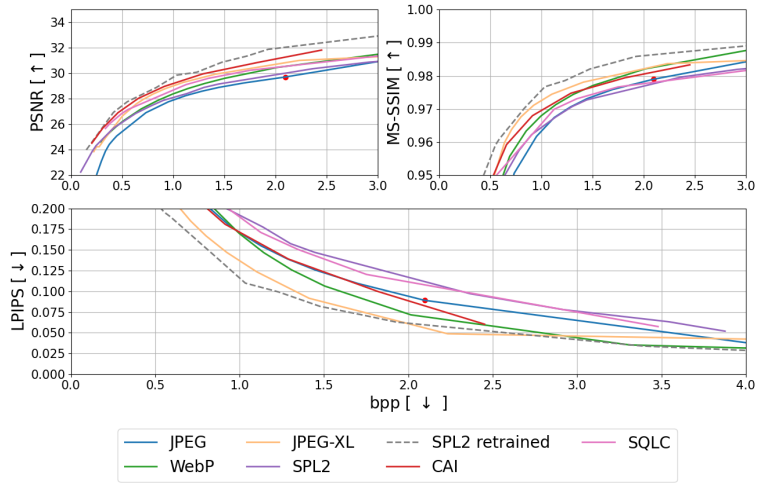

(a) Comparison of various lossy compression schemes on uncompressed data. Each line refers to one compression scheme.

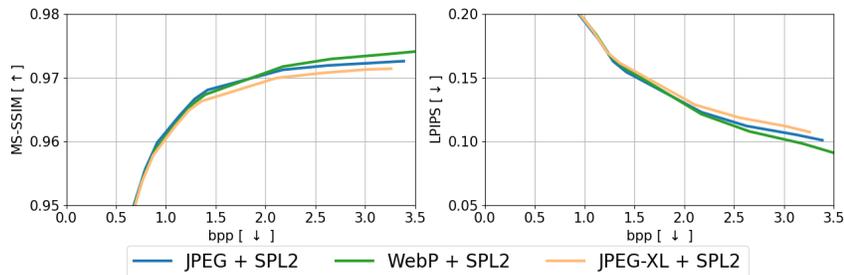

(b) A visualization of combined compression with JPEG, JPEG-XL and WebP with *SPL2* applied on top.

Fig. 4: Image quality metrics for compressing uncompressed *RMS* data. Figure 4a shows the results of methods without finetuning strategies, while fig. 4b shows the combined *SPL2* methods.

## 4.2 Downstream Task Image Quality

Segmentation Task In Figure 5a, we show qualitative results of the segmentation performance on different compression schemes. It can be seen that especially on small and connecting cell nuclei, the segmentation performance is heavily dependent on the compression scheme. Additionally in Figure 5b, we show the quantitative segmentation accuracy with the DICE metric. We report the averaged accuracy during five-fold cross-validation, which was performed with the nnU-Net framework. The Figure shows the compression ratio along the x-axis and the Dice score along the y-axis. Each colored curve refers to the respective accuracy for different compression schemes. For visualization purposes, we show a logarithmic x-scale with the basis 3.

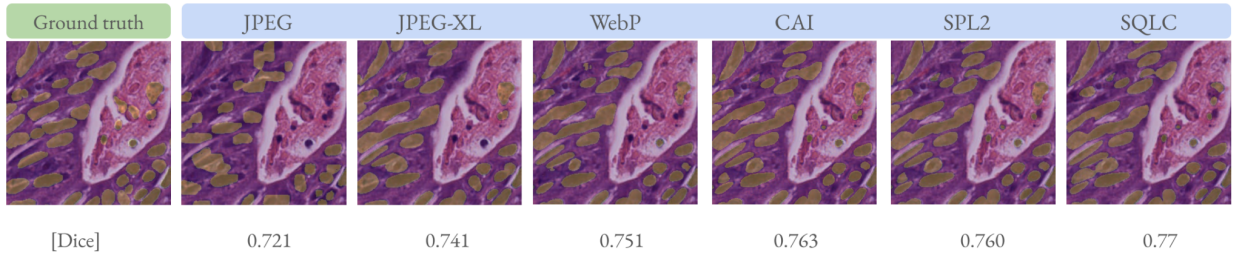

| Ground truth | JPEG | JPEG-XL | WebP | CAI | SPL2 | SQLC |
|---|---|---|---|---|---|---|
| [Dice] | 0.721 | 0.741 | 0.751 | 0.763 | 0.760 | 0.77 |

(a) Left: Ground truth image is shown with the corresponding segmentations. Right: Compressed images with various compression schemes and the predicted segmentations are shown as a yellow overlay. All images are compressed to a bitrate of 0.7 bpp.

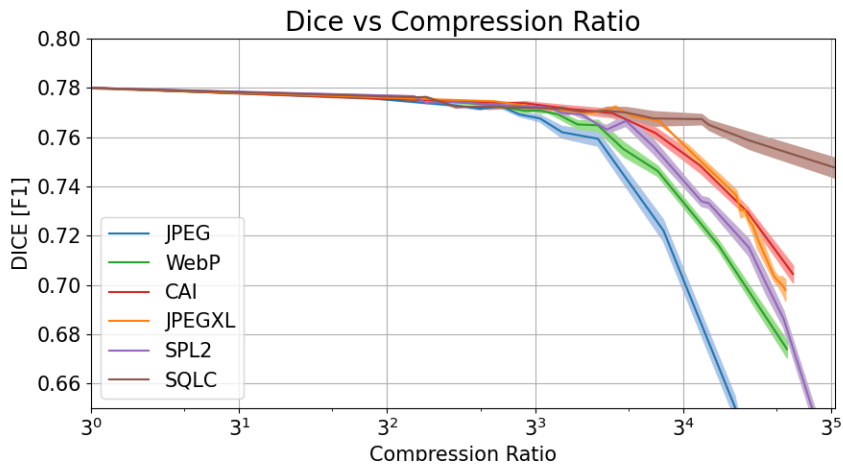

(b) Averaged segmentation accuracy during five-fold cross-validation. We report the averaged mean across all folds and show the standard deviation as a shaded area.

Fig. 5: Visualization of predicted segmentations along the numeric segmentation accuracy.

**Task Agnostic Metric** In Figure 6, the feature distances are shown for each compression scheme. The Figure shows that with increasing depth of the model, the mean distance between the original features and the compressed features is generally higher, while also more outliers can be observed. Comparing the best-performing compression scheme in the classification task from [26] and the segmentation task, it can be seen that this is also reflected by the on average highest mean similarity between the original image and the compressed image with *SQLC*.

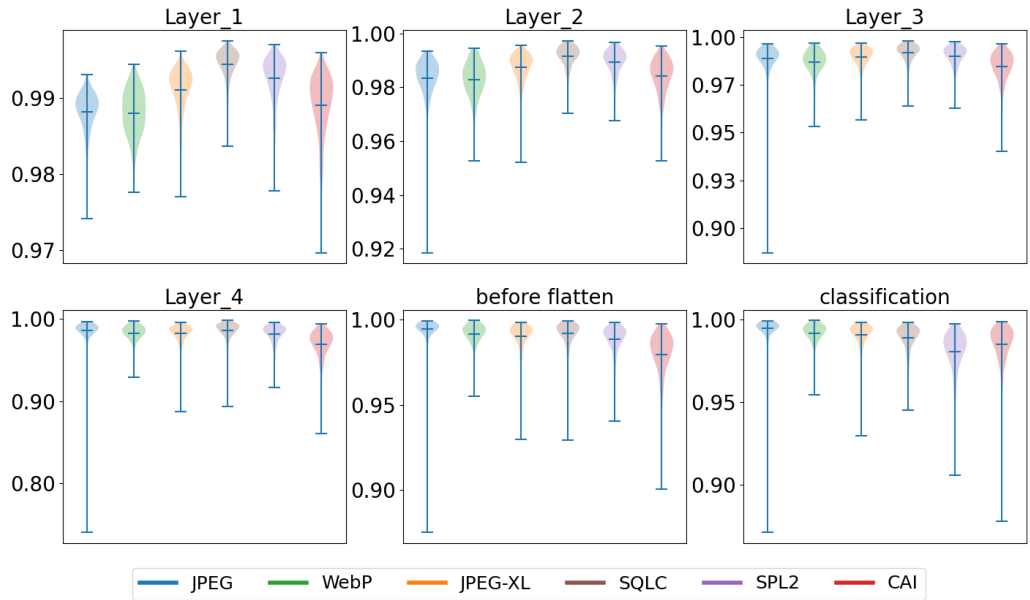

Fig. 6: Averaged cosine similarity between the original images and the compressed counterpart. Each violin refers to a different compression scheme and we extract features from different depths of the model. Along the y-axis the similarity measure is shown and each figure shows the similarity in a different layer.

## 4.3 Compression Time Evaluation

Our results in Figure 7 show the absolute en- and decoding times in a logarithmic scale on the x-axis for the test set for various bit-rates along the y-axis. Each bar represents here one of the compared schemes in this study. Our results show that dl-based compression schemes are slightly faster in encoding, compared to the conventional schemes. But decoding requires a tremendously longer time here. For more complex dl-schemes like *SQLC*, decoding times are also longer than for basic dl-schemes like *CAI*. In Figure 7, we the logarithmic time, which we measured in seconds. The absolute times to decode a slide from the RMS dataset took roughly 100 seconds with SQLC at bpp=0.1 and the encoding time was about 40 seconds at the same quality level.

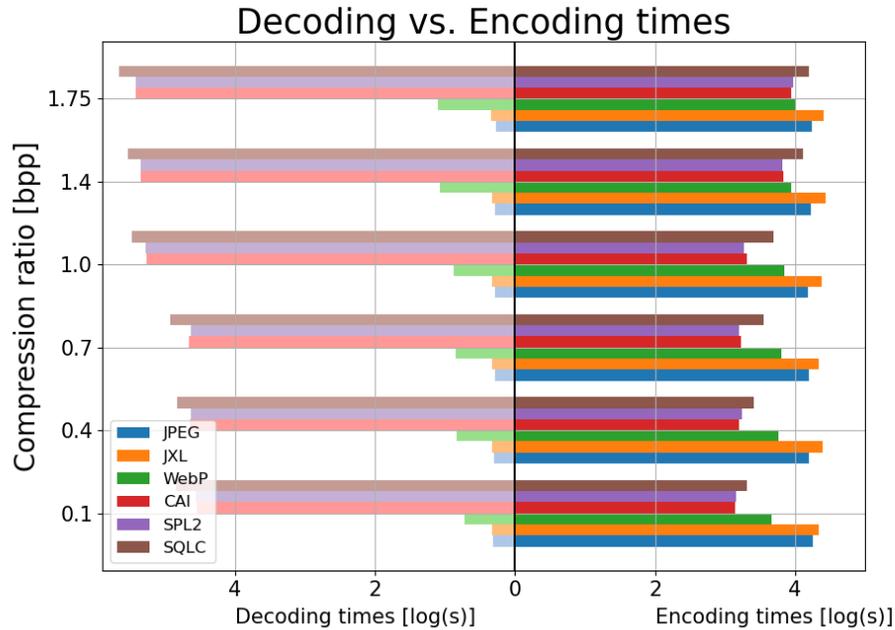

Fig. 7: De- vs. encoding times for the compared compression schemes on the test set. The Figure shows the measured times for various compression ratios.

## 5 Discussion

Deep learning-based lossy compression schemes as well as more recent conventional compression schemes outperform the lossy JPEG compression, which is the current state of the art in WSI acquisition, in various tasks. In this study, we compare the performance of different compression schemes in perceptual evaluations and in computational downstream tasks. In the first track of this study, we conduct experiments to evaluate (i) which compression scheme is best suited for recompressing data and (ii) which compression scheme is ideally suited for the initial compression. Our results show that for both tasks, a finetuned dl-based compression scheme that is tailored towards perceptual quality outperforms conventional and other dl-based schemes. However, our results also show that finetuned models fail to generalize between different datasets, which is shown by the performance of the *SPL2* that is trained on *JPEG80* data during inference on the *RMS* data. This shows that current training regimes are strongly biased towards the compression scheme in the training data and a pathological foundation model that generalizes across various compression schemes is yet to be established. In the second track of our study, we additionally evaluate compression schemes also for computational downstream algorithms. Here we assess (i) how do existing evaluations generalize across different tasks and (ii) task agnostic downstream metrics. Our results here also show that finetuning dl-compression schemes towards downstream performance again outperforms conventional compression schemes of schemes that are finetuned towards perceptual quality. Furthermore, we show that our proposed task-agnostic metric aligns very well with the performance in downstream tasks. This novel metric can help to accelerate research in this field, and to compare downstream performances for various compression schemes in a standardized way. In summary, our evaluations show that finetuning deep learning-based compression schemes achieve superior performance throughout various perceptual and downstream tasks. But this comes at the cost of significantly longer en- and decoding times of dl compression schemes compared to conventional schemes. The longer coding times paired with their current limitation to generalize across various datasets makes them not suited as novel clinical standard for WSI comression. Another limitation of neural compression schemes is the bias from their training data. Future models should carefully balance the distribution of training samples and the generalizability to novel unseen tissue types.

However, our study shows promising results for WebP, which in many experiments exhibited only marginally worse performances compared to the best-performing method, but with faster coding times. Further analyses will be conducted to also study the reliability of dl-based compression schemes. An open question here is how such methods perform when novel and unseen tissue types are compressed and how the downstream task performance is affected when certain classes of tissue are over- or underrepresented in the training data. In our study, we have considered various state-of-the-art lossy compression schemes, that perform real quantization. It is important to note that for our study we have only considered real lossy compression schemes that achieve file size reduction via quantization. We do not consider approaches with auto encoders, that solely perform dimensionality reduction or approaches that achieve file size reduction via removing diagnostically irrelevant parts of WSI like background [48]. The presented methods are all intended to compress tissue or foreground areas on images, that are diagnostically relevant with as much perceptual and downstream performance as possible. A promising alternative, which has not been evaluated in this study is the recently established JPEG-AI standard [49]. Initial experiments have shown great potential to address various shortcomings of the existing JPEG or JPEG-XL algorithms and to achieve competing results compared to other dl-based schemes. However, as of conducting these experiments, there was no publically available reference implementation. In future experiments, we plan to also consider the recent JPEG-AI standard in our experiments to further complete the evaluations. Additionally, further experiments with compression schemes in combination with clinical file formats are required.

## Conclusion

In this study we have evaluated the current state of the art lossy compression schemes, conventional as well as deep-learning based schemes on a broad range of pathology tasks. Our results show that conventional schemes often tend to generalize better across various tasks and datasets, while deep-learning based ones often exhibit a large bias from the training data. Finetuning specific compression schemes for single tasks and on specific datasets still yields the best performances. We also identified very large en- and decoding times especially for deep-learning based compression schemes. Future research should consider possibilities to accelerate deep-learning based compression to make it feasible for clinical evaluations. Optimizing Scanner hardware to support neural image compression schemes, as they support JPEG compression, could be a first step towards more suitable compression schemes in digital pathology beyond JPEG. Another important step are models that do not require time-consuming finetuning on datasets. Here large scale pretrained foundation models can help to mitigate dataset specific finetuning.

**Acknowledgement**
This work was supported in part by the DKTK Joint Funding UPGRADE, Project "Subtyping of pancreatic cancer based on radiographic and pathological features"(SUBPAN), by the Deutsche Forschungsgemeinschaft (DFG, German Research Foundation) under the grant 410981386 and by the Research Campus

M2OLIE, which was funded by the German Federal Ministry of Education and Research (BMBF) within the Framework "Forschungscampus: Public-private partnership for Innovations" under the funding code 13GW0388A.

**Declaration of Interest**
The authors declare no competing interests

**Declaration of generative AI and AI-assisted technologies in the writing process**
During the preparation of this work the authors used ChatGPT in order to improve readibility. After using this tool, the authors reviewed and edited the content as needed and take full responsibility for the content of the publication.

# Appendix

# Additional experiments on RMS data

Complementary to our experiments from section 4.1, we also performed initial experiments on downstream performances on the RMS data. With corresponding annotations and labels being made available recently in Imaging Data Commons [1 -3], also downstream performances can be evaluated on the RMS dataset. Existing studies in this field that investigate compression for dl-based downstream tasks mainly use already initially compressed data and evaluate how much additional compression is possible. Thus, an unbiased evaluation how much uncompressed data can be compressed without affecting performances is currently not available. Existing studies [4] suggest that compression up to a quality factor of 70 hardly effects any performance. With these initial additional experiments we aim to investigate this question with initially uncompressed data, which is not yet available. For our experiments, we again use the RMS dataset that consists of 96 subjects, that are divided into the two classes Embryonal rhabdomysarcoma and Alveolar rhabdomysarcoma. With this cohort, we trained a classification task with 52 Embryonal subjects and 44 Alveolar subjects. Existing work in this field already showed that generally classification in these two classes is possible with an accuracy of 0.85 ROC [2]. Here the authors focused on showing general classification capabilities on this task and omitted any experiments about compression schemes. Since the RMS dataset is originally uncompressed, this yields great potential for unbiased evaluations about the impact of compression schemes during downstream tasks. To evaluate it, we trained a 5-fold cross-validation multiple instance learning approach to classify on a slide level into the two classes. We used balanced folds and followed the common approach to use pre-trained feature extractors that extract embeddings for each patch in one slide and aggregate all embeddings to one lower dimensional bag of features per slide. For classifying each bag, we trained an attention-based Multiple Instance Learning model (abmil) [5]. For all our experiments, the abmil settings remained the same, where we used the Gated Attention model which we trained for 200 epochs with the Adam optimizer, a learning rate of 0.0005 and a weight decay of 0.0001. Throughout our experiments, we kept the abmil settings and splits fixed. Our experimental setting focused on the evaluation of lossy compression in downstream tasks. Thus we evaluated on different compression schemes during feature extraction. First, we used the openslide library [6] and extracted quadratic tiles with 224 pixels and a tissue threshold of 30% per tile. For each tile we computed embeddings with one of our compared feature extractors and aggregated all patch features of one slide to one bag. As feature extractor, we compared two different models: (i) a ResNet18 which was pretrained contrastively on the original RMS data and (ii) a ResNet18 which was pretrained contrastively on the RMS data, which we compressed with lossy JPEG and a quality factor of 80 to mimic pre-training scenarios which are common (most available pathology datasets that are used for pretraining are initially JPEG80 compressed). We refer to these feature extractors as **Contrastive** and **Contrastive_80.** We performed the contrastive pre-training ourself on the RMS data with the following augmentations: RandomResizedCrop, RandomHorizontalFlip, ColorJitter, RandomGrayscale and GaussianBlur. For contrastive pre-training, we used the framework from [7]. We kept the same training and validation splits that we also for the subsequent classification task. As an additional experiment, we also considered the UNI-foundation model from [8] as a universal foundation model for feature extraction. For all of our experiments, we used the same amount of patches and kept everything similar, except for the compression ratio that we applied on the patches during feature extraction. We synthetically compressed each patch to one of the JPEG compression ratios [original, 90, 80, 70], where original refers to no compression. In Figure 1, we show the classification accuracy [AUC] on the test set during 5 fold cross-validation and in table 2, we show the numerical results.

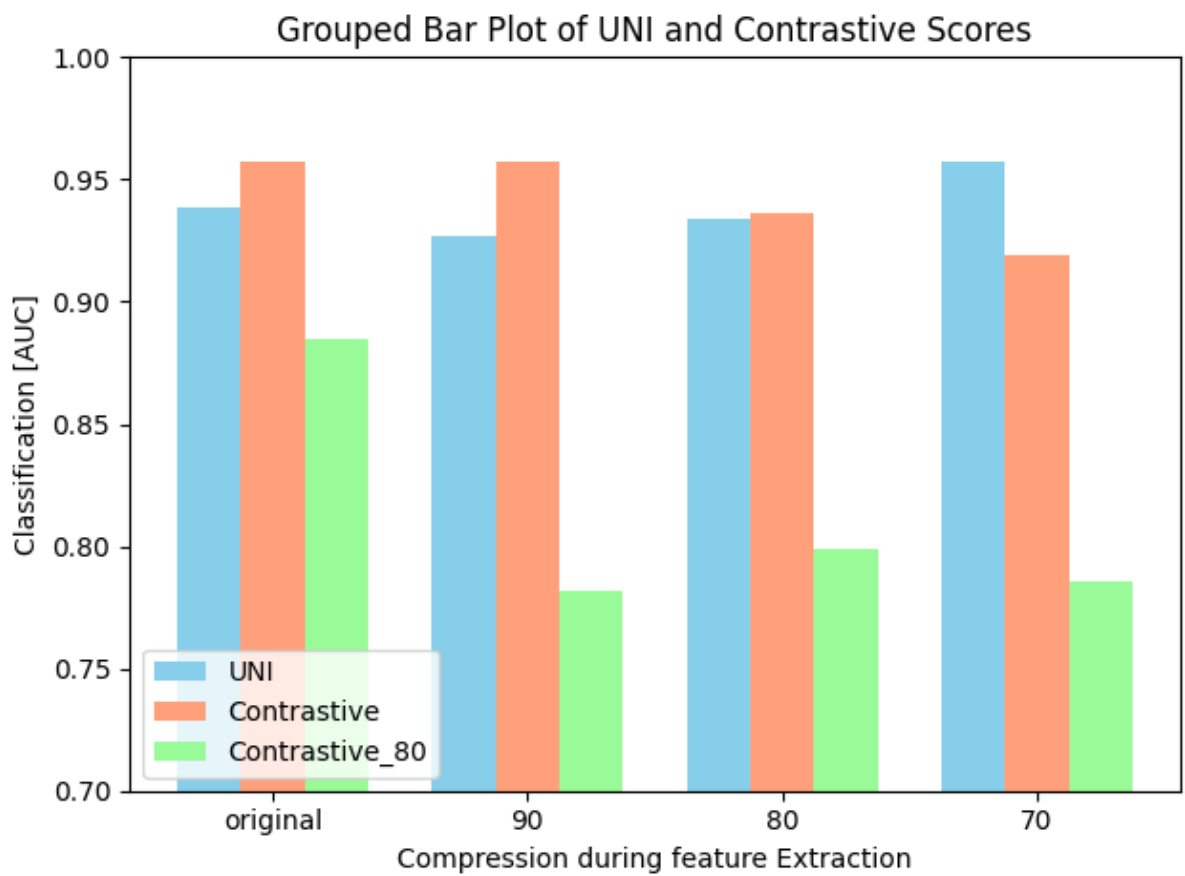

Figure1: Results of our classification experiments on the RMS data. Each bar shows the abmil performance with one feature extractor. On the x-axis we show the degree of compression that was applied during feature extraction and along the y-axis, we show the classification performance.

As our results indicate, for inference on the uncompressed RMS data, the model that was also trained on the original uncompressed RMS data performs best (***Contrastive*** = 0.95). With increasing compression during feature extraction of the ***Contrastive*** model, the performance decreases. UNI, the large scale foundation model achieves best performances with JPEG70 being applied during feature extraction. These results indicate that for UNI, applying artificial lossy compression can be beneficial for downstream performance when uncompressed data is used.

Table1: Numeric results of our experiments. In our experiments, Contrastive pre-training on the original uncompressed RMS data performed best. Applying large scale out of domain yields to a huge performance drop, as well as pretraining on lossy compressed data.

| Feature Extractors | Original | JPEG 90 | JPEG 80 | JPEG 70 |
|---|---|---|---|---|
| ***Contrastive*** | **0.9575** | **0.9575** | **0.9363** | 0.9191 |
| ***Contrastive_80*** | 0.7046 | 0.6016 | 0.6192 | 0.606 |
| UNI | 0.9383 | 0.9267 | 0.9343 | **0.9575** |

With our results we have shown that for classifying the RMS data into Aleveloar or Embryonal subtypes, it matters which compression is applied on the data that is used for pretraining feature extractors. In contrast to existing studies, we show that large amounts of performance are being lost when feature extractors are pretrained on compressed data, when pretraining on uncompressed data is also possible. This is shown by the comparison between the ***Contrastive*** model and the ***Contrastive_80*** model. As additional experiment, we also evlauated the robustness of foundation models against compression. Surprisingly, UNI together with abmil achieves best results, when JPEG70 compression is applied and not on the original data, which should be in theory of higher perceptual quality. Since the authors of [6] did not provide any information about the compression ratio that was present in their training data, it is hard to hypothesize why UNI works best for JPEG70. However the authors report that UNI was trained on an in-house hospital dataset. Thus we hypothesize that with JPEG70, UNI is applied in domain, since JPEG 80 or JPEG 70 are also common quality settings that are used in clinical pathology laboratories or publicly available data cohorts. However, our results also show a drastic performance drop when pretraining is done on JPEG80 and not with original data. Thus, our results suggest to stop applying lossy JPEG80 or 70 compression as a clinical standard. For many dl-based downstream tasks, a lot of performance is lost when a model is trained on JPEG80 data and not the original data. As a second recommendation, we also suggest to apply large scale foundation models also only at the same compression ratio they were pre-trained on. For various models this information is not available, but for most cases this would JPEG80 or 70. As a first step towards next experiments, we would suggest to investigate more recent compression schemes like WebP or JPEG-AI instead of lossy JPEG compression for such important tasks like digital pathology.

However, our results are only preliminary experiments and had several downsides. First, we did not finetune UNI on the specific RMS dataset. With a finetuned UNI encoder, UNI might even outperform ***Contrastive*** on the original data. And second, we did not address the dimensionality difference between the features extracted with UNI and the ***Contrastive*** model. UNI extracted twice as much features as the ResNets, but still could not outperform the finetuned ***Contrastive*** model. Future experiments should address this difference properly.

## References Appendix: